\documentclass[oneside,10pt]{article}

\title{\textbf{\Large{Adaptive Codes: A New Class of Non-standard Variable-length Codes}}}
\date{\normalsize{Drago\c s Trinc\u a\\
	Department of Computer Science \& Engineering\\University of Connecticut\\Storrs, CT 06269, USA\\
	\texttt{dtrinca@engr.uconn.edu}}}

\usepackage{amssymb}

\newcommand{\ac}[1]{c:\Sigma\times\Sigma^{\leq{#1}}\rightarrow\Delta^{+}}
\newcommand{\eac}{\overline{c}:\Sigma^{*}\rightarrow\Delta^{*}}
\newcommand{\sstring}[1]{\sigma_{1}\sigma_{2}\ldots\sigma_{#1}}
\newcommand{\ttup}[1]{\texttt{\textup{#1}}}

\newcounter{section_2}
\newtheorem{definition_2}{Definition}[section_2]
\newtheorem{remark_2}{Remark}[section_2]
\newtheorem{example_2}{Example}[section_2]
\newtheorem{theorem_2}{Theorem}[section_2]

\newcounter{section_3}
\newtheorem{definition_3}{Definition}[section_3]
\newtheorem{remark_3}{Remark}[section_3]
\newtheorem{example_3}{Example}[section_3]
\newtheorem{proposition_3}{Proposition}[section_3]

\newcounter{section_4}
\newtheorem{definition_4}{Definition}[section_4]

\newcounter{section_5}
\newtheorem{definition_5}{Definition}[section_5]
\newtheorem{remark_5}{Remark}[section_5]
\newtheorem{theorem_5}{Theorem}[section_5]

\begin{document}

\maketitle

\begin{abstract}
	We introduce a new class of non-standard variable-length
	codes, called \textit{adaptive codes}. This class of codes associates a variable-length
	codeword to the symbol being encoded depending on the previous symbols in the
	input data string. An efficient algorithm for constructing adaptive codes of order one
	is presented. Then, we introduce a natural generalization of adaptive codes, called
	\textit{GA codes}.
	
	{\bf Keywords:} adaptive mechanisms, compression rate, data compression, entropy, prefix codes, variable-length codes.
\end{abstract}

\section{Introduction}
	The theory of variable-length codes \cite{bp1} originated in concrete
	problems of information transmission. Especially by its language theoretic
	branch, the field has produced a great number of results, most of them with
	multiple applications in engineering and computer science.
	Intuitively, a \textit{variable-length code} is a set of strings such that any concatenation of these strings
	can be uniquely decoded. We introduce a new class of non-standard variable-length codes,
	called
	\textit{adaptive codes}, which associate a variable-length codeword to the
	symbol being encoded depending on the previous symbols in the input data
	string.
\newline\indent
	The paper is organized into six sections. After this introductory section,
	the definition of adaptive codes and several theoretical remarks 
	are given in Section 2, as well as some characterization results for 
	adaptive codes. The main results of this paper are presented in Section 3, 
	where we focus on designing an algorithm for constructing adaptive codes 
	of order one.
	In Section 4, we compute the entropy bounds for this algorithm.
	A natural generalization
	of adaptive codes is presented in Section 5. Finally, the last
	section contains a few concluding remarks.
\newline\indent
	Before\hspace{1pt} ending this introductory section, let us present
	some useful notation used throughout the paper \cite{rs1, as1}, and then review some basic concepts.
	We denote by $|S|$ the \textit{cardinality} of a set $S$; if $x$ is a string of
	finite length, then $|x|$ denotes the length of $x$.
	The \textit{empty string} is denoted by $\lambda$.
\newline\indent
	For an alphabet $\Sigma$,
	we denote by $\Sigma^{*}$ the set
	$\bigcup_{n=0}^{\infty}\Sigma^{n}$,
	and by $\Sigma^{+}$ the set
	$\bigcup_{n=1}^{\infty}\Sigma^{n}$,
	where $\Sigma^{0}$ is defined by $\{\lambda\}$.
	Let us denote by
	$\Sigma^{\leq n}$ the set
	$\bigcup_{i=0}^{n}\Sigma^{i}$
	and by $\Sigma^{\geq n}$ the set
	$\bigcup_{i=n}^{\infty}\Sigma^{i}$.
	Let $X$ be a finite and nonempty subset of $\Sigma^{+}$, and $w\in\Sigma^{+}$.
	A \textit{decomposition of w} over $X$ is any sequence of words
	$u_{1}, u_{2}, \ldots, u_{h}$ with $u_{i}\in X$, $1\leq i\leq h$, 
	such that $w=u_{1}u_{2}\ldots u_{h}$.
	A \textit{code} over $\Sigma$ is any nonempty set $C\subseteq\Sigma^{+}$ such
	that each word $w\in\Sigma^{+}$ has at most one decomposition over $C$.
	A \textit{prefix code} over $\Sigma$ is any code $C$ over $\Sigma$ such that
	no word in $C$ is proper prefix of another word in $C$.

\section{Adaptive Codes}
	In this section we introduce a new class of non-standard variable-length codes, called adaptive codes.
	These codes are based on adaptive mechanisms, that is, the variable-length codeword
	associated to the symbol being encoded depends on the previous symbols in the input data string.
\begin{definition_2}
	Let $\Sigma$ and $\Delta$ be two alphabets. A function
	$\ac{n}$, with $n\geq{1}$, is called an \textup{adaptive code of order n} if its unique
	homomorphic extension $\eac$ given by:
\begin{itemize}
\item $\overline{c}(\lambda)=\lambda$,
\item $\overline{c}(\sstring{m})=$
	$c(\sigma_{1},\lambda)$
	$c(\sigma_{2},\sigma_{1})$
	$\ldots$
	$c(\sigma_{n-1},\sstring{n-2})$
	\newline
	$c(\sigma_{n},\sstring{n-1})$
	$c(\sigma_{n+1},\sstring{n})$
	$c(\sigma_{n+2},\sigma_{2}\sigma_{3}\ldots\sigma_{n+1})$
	\newline
	$c(\sigma_{n+3},\sigma_{3}\sigma_{4}\ldots\sigma_{n+2})\ldots$
	$c(\sigma_{m},\sigma_{m-n}\sigma_{m-n+1}\ldots\sigma_{m-1})$,
\end{itemize}
for all $\sstring{m}\in\Sigma^{+}$, is injective.
\end{definition_2}
\hspace{0pt}\indent
	Let us take an example in order to better understand the adaptive mechanisms presented in the definition above.
\begin{example_2}
	Let $\Sigma=\{\ttup{a},\ttup{b}\}$, $\Delta=\{0,1\}$ be alphabets, and 
	$\ac{2}$ a function given by the table below.
\begin{center}
\textup{
\begin{tabular}{|c|c|c|c|c|c|c|c|}
\hline
$\Sigma\backslash\Sigma^{\leq{2}}$ 	& \ttup{a} 	& \ttup{b} 	& \ttup{aa} 	& \ttup{ab} 	& \ttup{ba} 	& \ttup{bb} 	& $\lambda$	\\ \hline
\ttup{a} 					& 0 	& 0 	& 0 	& 0 	& 1 	& 1 	& 0 		\\ \hline
\ttup{b}					& 1	& 1	& 1	& 1	& 0	& 0	& 1		\\ \hline
\end{tabular}}	
\end{center}
	One can easily verify that the function $\overline{c}$ is injective, and according to Definition 2.1,
	$c$ is
	an adaptive code of order two. Let $x=\ttup{abaa}\in\Sigma^{+}$. 
	Using the definition above, we encode $x$ by
$\overline{c}(x)=c(\ttup{a},\lambda)c(\ttup{b},\ttup{a})c(\ttup{a},\ttup{ab})c(\ttup{a},\ttup{ba})=0101$.
\end{example_2}
\hspace{0pt}\indent
	Let $\ac{n}$ be an adaptive code of order $n$, $n\geq{1}$. We denote by
	$C_{c, \sigma_{1}\sigma_{2}\ldots\sigma_{h}}$ the set
\begin{center}
	$\{c(\sigma,\sigma_{1}\sigma_{2}\ldots\sigma_{h}) \mid \sigma\in\Sigma\}$,
\end{center}
	for all $\sigma_{1}\sigma_{2}\ldots\sigma_{h}\in\Sigma^{\leq{n}}-\{\lambda\}$,
	and by $C_{c, \lambda}$ the set $\{c(\sigma,\lambda) \mid \sigma\in\Sigma\}$.
	We write $C_{\sigma_{1}\sigma_{2}\ldots\sigma_{h}}$ instead of
	$C_{c, \sigma_{1}\sigma_{2}\ldots\sigma_{h}}$, and $C_{\lambda}$ instead of $C_{c, \lambda}$
	whenever there is no confusion.
	
	If $w\in\Sigma^{+}$ then we denote by $w(i)$ the $i$-th symbol of $w$.
	In the rest of this paper we denote by ${\it AC}(\Sigma,\Delta,n)$ the set
\begin{center}
	$\{\ac{n} \mid$ $c$ is an adaptive code of order $n\}$.
\end{center}
\begin{theorem_2}
	Let $\Sigma$ and $\Delta$ be two alphabets, and let $\ac{n}$ be a function.
	If $C_{\sstring{h}}$ is a prefix code, for all $\sstring{h}\in\Sigma^{\leq{n}}$,
	then $c\in{{\it AC}(\Sigma,\Delta,n)}$.
\end{theorem_2}
\textbf{Proof}
\hspace{10pt}
	Let us assume that $C_{\sstring{h}}$ is prefix code, for all
	$\sstring{h}\in\Sigma^{\leq{n}}$, but $c\notin{{\it AC}(\Sigma,\Delta,n)}$.
	By Definition 2.1, the unique homomorphic extension of c, denoted by $\overline{c}$,
	is not injective.
	This implies that $\exists$ $u\sigma u', u\sigma'u''\in\Sigma^{+}$, with
	$\sigma,\sigma '\in\Sigma$ and $u,u',u''\in\Sigma^{*}$, such that
	$\sigma\neq\sigma'$ and
\begin{center}
	$(*)$ $ $ $\overline{c}(u\sigma u')=\overline{c}(u\sigma'u'')$.
\end{center}
	We can rewrite $(*)$ by
\begin{center}
	$(**)$ $ $ $\overline{c}(u)c(\sigma,P_{n}(u))\overline{c}(u')=$
	$\overline{c}(u)c(\sigma',P_{n}(u))\overline{c}(u'')$,
\end{center}
	where $P_{n}(u)$ is given by
\begin{displaymath}
P_{n}(u)=
\left\{ 	
	\begin{array}{ll}
	\lambda 			& \textrm{if $u=\lambda$,} \\
	u_{1}\ldots u_{q} 	& \textrm{if $u=u_{1}u_{2}\ldots u_{q}$ and $u_{1},u_{2},\ldots,u_{q}\in\Sigma$ and $q\leq{n}$,} \\
	u_{q-n+1}\ldots u_{q} 	& \textrm{if $u=u_{1}u_{2}\ldots u_{q}$ and $u_{1},u_{2},\ldots,u_{q}\in\Sigma$ and $q>n$.}
	\end{array} 
\right.
\end{displaymath}
	By hypothesis, $C_{P_{n}(u)}$ is a prefix code and
	$c(\sigma,P_{n}(u)),c(\sigma',P_{n}(u))\in{C_{P_{n}(u)}}$.
	Therefore, the set $\{c(\sigma,P_{n}(u)),c(\sigma',P_{n}(u))\}$ is a prefix code.
	But the equality $(**)$ can hold if and only if
	$\{c(\sigma,P_{n}(u)),c(\sigma',P_{n}(u))\}$ is not a prefix set.
	Hence, our assumption leads to a contradiction.
$\diamondsuit$
\begin{remark_2}
	The converse of Theorem 2.1 does not hold. We can prove this by taking a counter-example.
	Let us consider $\Sigma=\{\ttup{a},\ttup{b}\}$ and $\Delta=\{0,1\}$ two alphabets, and
	$\ac{2}$ a function given by the table below.
\begin{center}
\textup{
\begin{tabular}{|c|c|c|c|c|c|c|c|}
\hline
$\Sigma\backslash\Sigma^{\leq{2}}$ 	& \ttup{a} 	& \ttup{b} 	& \ttup{aa} 	& \ttup{ab} 	& \ttup{ba} 	& \ttup{bb} 	& $\lambda$	\\ \hline
\ttup{a} 					& 0 	& 0 	& 0 	& 0 	& 0 	& 0 	& 0 		\\ \hline
\ttup{b}					& 01	& 1	& 1	& 1	& 1	& 1	& 1		\\ \hline
\end{tabular}}
\end{center}
	One can verify that the unique homomorphic extension of $c$, denoted by
	$\overline{c}$, is injective. Therefore, we conclude that the function $c$ is an adaptive code
	of order two.
\end{remark_2}
\hspace{0pt}\indent
	Let $\Sigma$, $\Delta$, and ${\it Bool}=\{{\it True}, {\it False}\}$ be alphabets.
	We define the function
	${\it Prefix}:{\it AC}(\Sigma,\Delta,n)\rightarrow {\it Bool}$ by:
\begin{displaymath}
{\it Prefix}(c)=
\left\{ 	
	\begin{array}{ll}
	{\it True} 			& \textrm{if $C_{u}$ is a prefix code, for all $u\in\Sigma^{\leq{n}}$,} \\
	{\it False}		 	& \textrm{otherwise.}
	\end{array} 
\right.
\end{displaymath}
\hspace{10pt}
\newline\indent
	The function \textit{Prefix} can now be used to translate the hypothesis in Theorem 2.1:
	if $\ac{n}$ is a function satisfying ${\it Prefix}(c)={\it True}$, then we conclude that
	$c\in{{\it AC}(\Sigma,\Delta,n)}$.
\newline\indent
	Let $c\in{{\it AC}(\Sigma,\Delta,n)}$ be an adaptive code satisfying
	${\it Prefix}(c)={\it True}$. Then, the algorithm \textbf{Decoder} described below requires a linear time.
\newline
\setlength{\unitlength}{1pt}
\begin{picture}(340,240)
	\put(30,211){\textbf{Decoder}$(c,u)$}
	\put(30,199){input:\hspace{18pt}$c\in{{\it AC}(\Sigma,\Delta,n)}$ \textit{such that} ${\it Prefix}(c)={\it True}$ \textit{and}
		$u\in\Delta^{+}$;}
	\put(30,187){output:\hspace{11pt}$w\in\Sigma^{+}$ \textit{such that} $\overline{c}(w)=u$;}
	\put(30,175){begin}
	\put(15,163){1.\indent\indent
		$w:=\lambda$; $i:=1$; $Last:=\lambda$; $length:=|u|;$}
	\put(15,151){2.\indent\indent	while $i\leq{length}$ do}
	\put(15,139){\indent\indent\hspace{7.5pt}begin}
	\put(15,127){3.\indent\indent\indent\indent
		\textit{Let} $\sigma\in\Sigma$ \textit{be the unique symbol of $\Sigma$ with the property}}
	\put(15,115){\indent\indent\indent\hspace{19.4pt}
		\textit{that} $c(\sigma,Last)$ \textit{is prefix of} $u(i)\cdot u(i+1)\cdot\ldots\cdot u(length)$;}
	\put(15,103){4.\indent\indent\indent\indent$w:=w\cdot\sigma$;}
	\put(15,91){5.\indent\indent\indent\indent$i:=i+|c(\sigma,Last)|$;}
	\put(15,79){6.\indent\indent\indent\indent if $|Last|<n$}
	\put(15,67){7.\indent\indent\indent\indent then $Last:=Last\cdot\sigma$;}
	\put(15,55){8.\indent\indent\indent\indent else $ $
		$Last:=Last(|Last|-n+2)\cdot\ldots\cdot Last(|Last|)\cdot\sigma$;}
	\put(15,43){\indent\indent\hspace{7.5pt}end}
	\put(15,31){9.\indent\indent return $w$;}	
	\put(30,19){end}
	\put(8,227){\line(1,0){327}}
	\put(8,9){\line(1,0){327}}
	\put(8,227){\line(0,-1){218}}
	\put(335,227){\line(0,-1){218}}
\end{picture}
\begin{remark_2}
	In the third step of the algorithm given above, the symbol denoted by $\sigma$ is unique with that
	property due to the input restrictions.
\end{remark_2}
\begin{remark_2}
	One can easily verify that the while loop in algorithm \textbf{\textup{Decoder}}
	is iterated 
\begin{displaymath}
	|u|-\sum_{i=1}^{h}(|c(w_{i},P_{n}(w_{1}w_{2}\ldots w_{i-1}))|-1)
\end{displaymath}
	times, where $w=w_{1}w_{2}\ldots w_{h}$,
	and $P_{n}$ is the function given in Theorem 2.1.
\end{remark_2}

	In practice, we can use only adaptive codes satisfying the
	equality ${\it Prefix}(c)={\it True}$, since designing a decoding algorithm for the other case
	requires additional information and more complicated techniques. 

\section{Data Compression using Adaptive Codes}
	The construction of adaptive codes requires different approaches, depending on the structure of
	the input data strings. 
	In this section, we focus on data compression using adaptive codes of order one.
\begin{definition_3}
	Let $\Sigma$ be an alphabet and $w=w_{1}w_{2}\ldots w_{h}\in\Sigma^{\geq{2}}$,
	with $w_{i}\in\Sigma$, for all $i\in\{1,2,\ldots,h\}$. A subword $uu$ of $w$, with $u\in\Sigma$, is called
	a \textup{pair} of w.
\end{definition_3}
\begin{remark_3}
	Let $\Sigma$ be an alphabet and $w=w_{1}w_{2}\ldots w_{h}\in\Sigma^{\geq{2}}$.
	It is useful to consider the following notations:
\begin{enumerate}
\item ${\it Pairs}(w)=\{i \mid 1\le i\le |w|-1, w_i=w_{i+1}\}$,
\item ${\it NRpairs}(w)=|{\it Pairs}(w)|$,
\item ${\it Prate}(w)=\frac{{\it NRpairs}(w)}{|w|}$.
\end{enumerate}
\end{remark_3}
\hspace{10pt}
	The main goal of this section is to design an algorithm for constructing
	adaptive codes of order one, under the assumption that the input data strings
	have a large number of pairs.
\newline\indent
	Let $\Sigma=\{\sigma_{1},\sigma_{2},\ldots,\sigma_{h}\}$ and $\Delta=\{0,1\}$
	be alphabets, $c\in{{\it AC}(\Sigma,\Delta,1)}$ an adaptive code of order one, and $w\in\Sigma^{+}$.
	We denote by $A_{c}$ the matrix given by:
\begin{center}
	$
	A_{c} =
	\left(\begin{array}{ccccc}
		c(\sigma_{1},\sigma_{1}) & c(\sigma_{1},\sigma_{2}) & \ldots & c(\sigma_{1},\sigma_{h}) & c(\sigma_{1},\lambda) \\
		c(\sigma_{2},\sigma_{1}) & c(\sigma_{2},\sigma_{2}) & \ldots & c(\sigma_{2},\sigma_{h}) & c(\sigma_{2},\lambda) \\
						 &				    &	\ldots &				    &				    \\											
		c(\sigma_{h},\sigma_{1}) & c(\sigma_{h},\sigma_{2}) & \ldots & c(\sigma_{h},\sigma_{h}) & c(\sigma_{h},\lambda) 
		\end{array} \right)
	$.
\end{center}
\hspace{10pt}
	Let us denote by \textbf{Huffman}$({\it EF}(w),n)$ the well-known Huffman's algorithm \cite{ds1},
	where $n\geq{1}$, and ${\it EF}(w)$ is the matrix given below.
\begin{center}
	$
	{\it EF}(w) =
	\left(\begin{array}{ccccc}
		\sigma_{1}		& \sigma_{2}	& \ldots		& \sigma_{n}	\\
		f(\sigma_{1},w)	& f(\sigma_{2},w)	& \ldots	 	& f(\sigma_{n},w) 
		\end{array} \right)
	$.
\end{center}
\hspace{10pt}
	We assume that the first row of the matrix ${\it EF}(w)$ contains the symbols which
	are being encoded, while the second row contains their frequencies, that is, $f(\sigma_{i})$
	is the frequency of the symbol $\sigma_{i}$ in $w$.
\newline\indent
 	Also, we assume that \textbf{Huffman}$({\it EF}(w),n)$ is the matrix given by
\begin{center}
	$
	\textbf{Huffman}({\it EF}(w),n) =
	\left(\begin{array}{ccccc}
		H(\sigma_{1},w)		& H(\sigma_{2},w)	& \ldots		& H(\sigma_{n},w)	
		\end{array} \right)
	$
\end{center}
	where $H(\sigma_{i},w)$ is the codeword associated to the symbol $\sigma_{i}$
	by Huffman's algorithm.
	The algorithm \textbf{Builder} described further on takes linear time, and constructs
	an adaptive code of order one satisfying ${\it Prefix}(c)={\it True}$.
\begin{proposition_3}
	Let $c:\Sigma\times\Sigma^{\leq{1}}\rightarrow\{0,1\}^{+}$ be a function
	given by the matrix \textbf{Builder}$(c)$. Then, $c\in{{\it AC}(\Sigma,\{0,1\},1)}$ and
	${\it Prefix}(c)={\it True}$.
\end{proposition_3}
\textbf{Proof}
\hspace{10pt}
	Applying the algorithm \textbf{Builder} to the function $c$, one can easily verify
	that ${\it Prefix}(c)={\it True}$. Therefore, according to \textbf{Theorem 2.1},
	$c$ is an adaptive code of order one, that is, $c\in{{\it AC}(\Sigma,\{0,1\},1)}$.
$\diamondsuit$
\newpage
\setlength{\unitlength}{1pt}
\begin{picture}(340,282)
	\put(15,268){\textbf{Builder}(c)}
	\put(15,256){input:\hspace{18pt}$c:\Sigma\times\Sigma^{\leq{1}}\rightarrow\{0,1\}^{+}$,
		$\Sigma=\{\sigma_{1},\sigma_{2},\ldots,\sigma_{h}\}$;}
	\put(15,244){output:\hspace{11pt}$A_{c}$ \textit{such that} $c\in{{\it AC}(\Sigma,\{0,1\},1)}$
		\textit{and} ${\it Prefix}(c)={\it True}$;}
	\put(15,232){begin}
	\put(0,220){$1.$\indent\indent for $i:=1$ to $h$ do $A_{c}(i,i):=0$;}
	\put(0,198){2.\indent\indent
		$E:=	\left(
			\begin{array}{cccc}
			\sigma_{2} 	& \sigma_{3} 	& \ldots 	& \sigma_{h} \\
				0	&	0		& \ldots	&	0	 
			\end{array}
			\right)
		$;}
	\put(0,176){3.\indent\indent $X:=$\hspace{3pt}\textbf{Huffman}$(E,h-1)$;}
	\put(0,164){4.\indent\indent for $i:=2$ to $h$ do}
	\put(0,152){\indent\indent\hspace{4pt} begin}
	\put(0,140){5.\indent\indent\indent\indent
		$A_{c}(1,i):=1\cdot X(1,i-1)$;}
	\put(0,128){6.\indent\indent\indent\indent$X(1,i-1):=1\cdot X(1,i-1)$;}
	\put(0,116){7.\indent\indent\indent\indent
		$A_{c}(i,1):=X(1,i-1)$;}
	\put(0,104){\indent\indent\hspace{4pt} end}
	\put(0,92){8.\indent\indent for $j:=2$ to $h$ do}
	\put(0,80){\indent\indent\hspace{4pt} begin}
	\put(0,68){9.\indent\indent\indent\indent for $i:=2$ to $j-1$ do $A_{c}(i,j):=X(1,i-1)$;}
	\put(-5,56){10.\indent\indent\indent\indent for $i:=j+1$ to $h$ do $A_{c}(i,j):=X(1,i-1)$;}
	\put(0,44){\indent\indent\hspace{4pt} end}
	\put(-5,32){11.\indent\indent for $i:=1$ to $h$ do $A_{c}(i,h+1):=A_{c}(1,i)$;}
	\put(-5,20){12.\indent\indent return $A_{c}$;}
	\put(15,8){end}
	\put(-8,282){\line(1,0){330}}
	\put(-8,2){\line(1,0){330}}
	\put(-8,282){\line(0,-1){280}}
	\put(322,282){\line(0,-1){280}}
\end{picture}
\hspace{0pt}
\begin{example_3}
	Let $c:\{\ttup{a},\ttup{b},\ttup{c}\}\times\{\ttup{a},\ttup{b},\ttup{c}\}^{\leq{1}}\rightarrow\{0,1\}^{+}$ be a function.
	One can verify that $A_{c}$ is the matrix given below.
\begin{center}
	$A_{c}=\left(
		\begin{array}{cccc}
			0 	& 	10 	& 	11 	& 	0 		\\
			11	&	0	& 	10	&	10	 	\\
			10	&	11	& 	0	&	11	 
		\end{array}
		\right)
	$.
\end{center}
\hspace{10pt}
	Let $w=\ttup{abbbcabccaabccabbcba}$ be an input data string. It is easy to verify that
	${\it Pairs}(w)=\{2,3,8,10,13,16\}$, ${\it NRpairs}(w)=6$, and ${\it Prate}(w)=0.3$.
\newline\indent
	Encoding the string $w$ by $c$ requires the computation of $\overline{c}(w)$.
	Using Definition 2.1, we get that $|\overline{c}(w)|=33$.
\newline\indent
	Let us apply Huffman's algorithm to the data string $w$ in order to make
	a comparison between the results. If we denote by ${\it Huffman}(w)$ the codeword
	associated to $w$ by Huffman's algorithm, we get that
	$|Huffman(w)|=32$. An even better result can be obtained when the input data
	string has a larger number of pairs, as shown in the following example.
\end{example_3}
\begin{example_3}
	Let $c:\{\ttup{a},\ttup{b},\ttup{c}\}\times\{\ttup{a},\ttup{b},\ttup{c}\}^{\leq{1}}\rightarrow\{0,1\}^{+}$ be an
	adaptive code of order one given as in the previous example, and
	$w=\ttup{abbbccbccaabccaaacba}$ an input data string.
	One can verify that ${\it Pairs}(w)=\{2,3,5,8,10,13,15,16\}$,
	${\it NRpairs}(w)=8$, ${\it Prate}(w)=0.4$, and $|\overline{c}(w)|=31$.
	Encoding the string $w$ by Huffman's algorithm, we get that $|{\it Huffman}(w)|=34$. 
\end{example_3}
\hspace{10pt}
	The results obtained in the previous examples are summarized in the table below,
	which shows that we get substantial improvements for input data strings having a larger number of pairs.
\begin{center}
\textup{
\begin{tabular}{|c|c|c|c|c|c|c|c|}
\hline
$w$ 				& ${\it NRpairs}(w)$ 	& ${\it Prate}(w)$	&   $|\overline{c}(w)|$ 	& $|{\it Huffman}(w)|$ 		\\ \hline
$\ttup{abbbcabccaabccabbcba}$ 	& 6	 		& 0.3 		&   33 				& 32 				\\ \hline
$\ttup{abbbccbccaabccaaacba}$	& 8			& 0.4			&   31				& 34				\\ \hline
\end{tabular}}	
\end{center}

\section{Builder: Entropy Bounds}
	In this section, we focus on computing the entropy bounds for the algorithm described in section 3.
	Given that our algorithm is based on Huffman's algorithm, let us first recall
	the entropy bounds for Huffman codes.
\begin{definition_4}
	Let $\Sigma$ be an alphabet, $x$ a data string of length $n$ over $\Sigma$
	and $k$ the length of the encoder output, when the input is $x$.
	The \textup{compression rate}, denoted by $R(x)$, is defined by
\begin{displaymath}
	R(x)=\frac{k}{n}.
\end{displaymath}
\end{definition_4}
\hspace{10pt}
	Let $R(x)$ be the compression rate in codebits per datasample, computed after
	encoding the data string $x$ by the Huffman algorithm. One can obtain
	upper and lower bounds on $R(x)$ before encoding the data string $x$ by computing
	the entropy denoted by $H(x)$.
	Let $x$ be a data string of length $n$, $(F_{1},F_{2},\ldots,F_{h})$
	the vector of frequencies of the symbols in $x$ and $k$ the length of
	the encoder output. The entropy $H(x)$ of $x$ is defined by
\begin{displaymath}
	H(x)=\frac{1}{n}\sum_{i=1}^{h}F_{i}\log_{2}(\frac{n}{F_{i}}).
\end{displaymath}
\hspace{10pt}
	Let $L_{i}$ be the length of the codeword associated to the symbol with the
	frequency $F_{i}$ by the Huffman algorithm, $1\leq{i}\leq{h}$. Then,
	the compression rate $R(x)$ can be re-written by
\begin{displaymath}
	R(x)=\frac{1}{n}\sum_{i=1}^{h}F_{i}L_{i}.
\end{displaymath}
\hspace{10pt}
	If we relate the entropy $H(x)$ to the compression rate $R(x)$, we obtain
	the following inequalities:
\begin{displaymath}
	H(x)\leq{R(x)}\leq{H(x)+1}.
\end{displaymath}
\hspace{10pt}
	Let $\Sigma=\{\sigma_{1},\sigma_{2},\ldots,\sigma_{t}\}$ be an alphabet
	and $c:\Sigma\times\Sigma^{\leq{1}}\rightarrow\{0,1\}^{+}$ an adaptive
	code of order one constructed as shown in section 3.
	Also, consider $w=w_{1}w_{2}\ldots{w_{s}}\in{\Sigma^{+}}$,
	$w_{i}\in\Sigma$, $1\leq{i}\leq{s}$, and $p$ the number of symbols occurring in $w$.
\newline\indent
	We denote by $R_{A}(w)$ the compression rate obtained when encoding the string $w$
	by $\overline{c}$ and by $H_{A}(w)$ the entropy of $w$.
	It is useful to consider the following notations:
\begin{enumerate}
\item ${\it EH}(w)=\{i$ $\mid$ $2\leq{i}\leq{s}$ and $w_{i}\neq{w_{i-1}}\}$,
\item ${\it LNotHuffman}(w)=|c(w_{1},\lambda)|+\sum_{i\in{{\it Pairs}(w)}}|c(w_{i+1},w_{i})|$,
\item ${\it LHuffman}(w)$ is the entropy of $w_{j_{1}}w_{j_{2}}\ldots{w_{j_{r}}}$,
	$j_{k}\in{{\it EH}(w)}$, $1\leq{k}\leq{r}$,
\item $H_{A}(w)={\it LNotHuffman}(w)+{\it LHuffman}(w)$.
\end{enumerate}
\hspace{10pt}
	It is easy to verify that ${\it LNotHuffman}(w)={\it NRPairs}(w)+|c(w_{1},\lambda)|$.
	Using the notation above, we get that
\begin{displaymath}
	{\it LHuffman}(w)=\sum_{i\in{{\it EH}(w)}}\{ \frac{1}{N(w_{i})}\sum_{q\in{{\it Prev}(w_{i})}}[F_{q}(w_{i})(1+\log_{2}\frac{N(w_{i})}{F_{q}(w_{i})})]\},
\end{displaymath}
where
\begin{itemize}
\item $N(w_{i})=|\{j$ $\mid$ $j\in{{\it EH}(w)}$ and $w_{j}={w_{i}}\}|$,
\item ${\it Prev}(w_{i})=\{j$ $\mid$ $j+1\in{{\it EH}(w)}$ and $w_{j+1}=w_{i}\}$,
\item $F_{q}(w_{i})=|\{j$ $\mid$ $j\in{{\it EH}(w)}$ and $w_{j}=w_{i}$ and $w_{j-1}=w_{q}\}|$,
	$q\in{{\it Prev}(w_{i})}$.
\end{itemize}
\hspace{10pt}
	Finally, we can relate the entropy $H_{A}(w)$ to the compression rate $R_{A}(w)$
	by the following inequalities:
\begin{displaymath}
	H_{A}(w)\leq{R_{A}(w)}\leq{H_{A}(w)}+1,
\end{displaymath}
where $R_{A}(w)$ is given by
\begin{displaymath}
	R_{A}(w)=\frac{|c(w_{1},\lambda)c(w_{2},w_{1})\ldots c(w_{s},w_{s-1})|}{s}.
\end{displaymath}

\section{GA Codes}
	In this section, we introduce a natural generalization of adaptive codes (of any order),
	called \textit{GA codes} (\textbf{G}eneralized \textbf{A}daptive codes).
	Theorem 5.1 proves that adaptive codes are particular cases of GA codes.
\begin{definition_5}
	Let $\Sigma$ and $\Delta$ be two alphabets and
	$F:N^{*}\times\Sigma^{+}\rightarrow\Sigma^{*}$ a function, where $N$ is the set
	of natural numbers, and $N^{*}=N-\{0\}$. A function
	$c_{F}:\Sigma\times\Sigma^{*}\rightarrow\Delta^{+}$ is called
	a \textup{GA code} if its unique homomorphic extension
	$\overline{c_{F}}:\Sigma^{*}\rightarrow\Delta^{*}$ given by
\begin{itemize}
\item $\overline{c_{F}}(\lambda)=\lambda$,
\item $\overline{c_{F}}(\sstring{m})=c_{F}(\sigma_{1},F(1,\sstring{m}))\ldots c_{F}(\sigma_{m},F(m,\sstring{m}))$,
\end{itemize}
	for all $\sstring{m}\in\Sigma^{+}$, is injective.
\end{definition_5}
\begin{remark_5}
	The function $F$ in Definition 5.1 is called the \textup{adaptive function}
	of the GA code $c_{F}$.
	Clearly, a GA code $c_{F}$ can be constructed if its adaptive function $F$
	is already constructed.
\end{remark_5}
\begin{remark_5}
	Let $\Sigma$ and $\Delta$ be two alphabets. We denote by $GAC(\Sigma,\Delta)$ the set
	$\{c_{F}:\Sigma\times\Sigma^{*}\rightarrow\Delta^{+}$ $\mid$ $c_{F}$ is a GA code$\}$.
\end{remark_5}
\begin{theorem_5}
	Let $\Sigma$ and $\Delta$ be alphabets. Then,
	$AC(\Sigma,\Delta,n)\subset{GAC(\Sigma,\Delta)}$,
	for all $n\geq{1}$.
\end{theorem_5}
\textbf{Proof}
\hspace{10pt}
	Let $c_{F}\in{AC(\Sigma,\Delta,n)}$ be an adaptive code of order $n$, $n\geq{1}$,
	and $F:N^{*}\times\Sigma^{+}\rightarrow\Sigma^{*}$ a function given by:
\begin{displaymath}
F(i,\sstring{m})=
\left\{ 	
	\begin{array}{ll}
	\lambda 			
					& \textrm{if $i=1$ or $i>m$,}		 				\\
	\sstring{i-1}		
					& \textrm{if $2\leq{i}\leq{m}$ and $2\leq{i}\leq{n+1}$,} 	\\
	\sigma_{i-n}\sigma_{i-n+1}\ldots\sigma_{i-1} 	
					& \textrm{if $2\leq{i}\leq{m}$ and $i>n+1$,}		
	\end{array} 
\right.
\end{displaymath}
	for all $i\geq{1}$ and $\sstring{m}\in{\Sigma^{+}}$.
	One can verify that $|F(i,\sstring{m})|\leq{n}$, for all 
	$i\geq{1}$ and $\sstring{m}\in{\Sigma^{+}}$.
	According to Definition 2.1, the function $\overline{c_{F}}$ is given by:
\begin{itemize}
\item $\overline{c_{F}}(\lambda)=\lambda$,
\item $\overline{c_{F}}(\sstring{m})=$
	$c_{F}(\sigma_{1},\lambda)$
	$c_{F}(\sigma_{2},\sigma_{1})$
	$\ldots$
	$c_{F}(\sigma_{n-1},\sstring{n-2})$
	\newline
	$c_{F}(\sigma_{n},\sstring{n-1})$
	$c_{F}(\sigma_{n+1},\sstring{n})$
	$c_{F}(\sigma_{n+2},\sigma_{2}\sigma_{3}\ldots\sigma_{n+1})$
	\newline
	$c_{F}(\sigma_{n+3},\sigma_{3}\sigma_{4}\ldots\sigma_{n+2})\ldots$
	$c_{F}(\sigma_{m},\sigma_{m-n}\sigma_{m-n+1}\ldots\sigma_{m-1})$,
\end{itemize}
	for all $\sstring{m}\in\Sigma^{+}$.
	It is easy to remark that
\begin{itemize}
\item	$\overline{c_{F}}(\sstring{m})=c_{F}(\sigma_{1},F(1,\sstring{m}))\ldots c_{F}(\sigma_{m},F(m,\sstring{m}))$,
\end{itemize}
	for all $\sstring{m}\in\Sigma^{+}$, which proves the theorem.
$\diamondsuit$

\section{Conclusions and Future Work}
	We introduced a new class of non-standard variable-length codes,
	called adaptive codes, which associate
	a variable-length codeword to the symbol being encoded depending on the previous
	symbols in the input data string. The main results of this paper are presented
	in Section 3, where we have shown that if an input data string $x$ has a significant number of pairs,
	then a good compression rate is achieved when encoding $x$ by adaptive codes
	of order one.
\newline\indent
	In a further paper devoted to adaptive codes, we intend to extend the algorithm
	\textbf{Builder} to adaptive codes of any order.

\small{
}

\end{document}